\documentclass[conference]{IEEEtran}
\usepackage{graphicx} 
\usepackage{epstopdf}
\usepackage{subfigure}
\usepackage{stfloats}
\usepackage{algorithm, algorithmic}
\usepackage{diagbox}
\usepackage{multirow}
\usepackage{mathtools}
\usepackage{setspace}  
\usepackage{threeparttable}
\usepackage{textcomp,booktabs}
\usepackage[usenames,dvipsnames]{color}
\usepackage{colortbl}
\usepackage{indentfirst}
\usepackage{cite}
\usepackage{amsmath,amssymb,amsfonts}
\usepackage{algorithmic}
\usepackage{textcomp}
\usepackage{xcolor}
\usepackage{mathtools}
\usepackage{amsthm}
\usepackage{color}
\usepackage{enumitem}
\usepackage[hang,flushmargin]{footmisc}

\newtheorem{lemma}{\bf Lemma}

\theoremstyle{definition}

\begin{document}

%title, author, and authors' information
\title{Balancing Latency and Model Accuracy for Fluid Antenna-Assisted LM-Embedded MIMO Network}

\author{
    \IEEEauthorblockN{Yichen Jin\IEEEauthorrefmark{1}, Zongze Li\IEEEauthorrefmark{2}, Zeyi Ren\IEEEauthorrefmark{1}, Qingfeng Lin\IEEEauthorrefmark{1}, and Yik-Chung Wu\IEEEauthorrefmark{1}}
    \IEEEauthorblockA{\IEEEauthorrefmark{1}Department of Electrical and Electronic Engineering, The University of Hong Kong, Hong Kong\\ \IEEEauthorrefmark{2}Peng Cheng Laboratory, Shenzhen 518038, China}

    \IEEEauthorblockA{Emails: \{u3589542, renzeyi, qflin, ycwu\}@eee.hku.hk, lizz@pcl.ac.cn}
}

% \author{Yichen Jin, Zongze Li, Zeyi Ren, Qingfeng Lin, and Yik-Chung Wu
% \thanks{\quad Y. Jin, Z. Ren, Q. Lin, and Y.-C. Wu are with the Department of Electrical and Electronic Engineering, The University of Hong Kong, Hong Kong (e-mail: u3589542@eee.hku.hk, qflin@eee.hku.hk, zyren@eee.hku.hk, ycwu@eee.hku.hk).}
% \thanks{\quad Z. Li is with Peng Cheng Laboratory, Shenzhen, Guangdong 518055, China (e-mail: lizz@pcl.ac.cn).}
% }

\maketitle

%abstract
\begin{abstract}
% 引出LM，说明LM带来的问题（不能本地处理大量数据，需要边缘服务器），引出量化，再说量化不能完全解决问题，因为过于激进的量化会导致模型精度过低。于是，引入FA，通过提升信息传输速度来提升基于边缘网络的LM推理的质量。本文提出了一个什么样的系统，为了进行资源分配，而构造了什么问题，怎么解决。实验展示。

This paper addresses the challenge of large model (LM)-embedded wireless network for handling the trade-off problem of model accuracy and network latency. To guarantee a high-quality of users' service, the network latency should be minimized while maintaining an acceptable inference accuracy.  
To meet this requirement, LM quantization is proposed to reduce the latency. 
However, the excessive quantization may destroy the accuracy of LM inference. To this end, a promising fluid antenna (FA) technology is investigated for enhancing the transmission capacity, leading to a lower network latency in the LM-embedded multiple-input multiple-output (MIMO) network. 
To design the FA-assisted LM-embedded network with the lower latency and higher accuracy requirements, the latency and peak signal to-noise ratio (PSNR) are considered in the objective function.
Then, an efficient optimization algorithm is proposed under the block coordinate descent framework. Simulation results are provided to show the convergence behavior of the proposed algorithm, and the performance gains from the proposed FA-assisted LM-embedded network over the other benchmark networks in terms of network latency and PSNR.
\end{abstract}


%intro
\section{Introduction}

%第一段介绍LM，说LM的好处和问题，根据问题说LM的部署要基于边缘系统，而且推理延迟（包括传输和计算延迟）很重要。从‘为了降低延迟’引出模型量化，并说不能过分量化，这可能会严重降低推理精度，所以只是量化一个技术还是不能用有限的资源满足用户对推理延迟和精确度的综合要求

In the future, more and more wireless applications, such as autonomous driving, will be implemented based on large models (LMs). Although the integration of LM in 6G networks offers significant benefits, the LM requires substantial computational resources, which is challenging for resource-constrained and latency-sensitive wireless users. As a result, the LM is usually implemented on the edge or cloud servers. However, with the explosive growth of data in LM for offloading and training at servers, the network latency, including the offloading latency and inference latency, has significantly increased. Hence, it is crucial to ensure the network latency within the tolerance level for the LM-embedded wireless network~\cite{wohaoxianmu_LM}.

The traditional methods to reduce network latency rely on model quantization~\cite{Survey_quantize_NN_inference}, which enables the less storage space and the smaller volume of data offloading. However, the aggressive quantization may lead to a significant degradation in model accuracy, adversely affecting the users' experience. 
In balancing between the model accuracy and network latency, the fluid antenna (FA) is emerging as one of the most promising tools to reduce the network latency of wireless systems without sacrificing the model accuracy. 
To be specific, the FA provides the higher spatial degrees of freedom (DoFs) by adjusting the physical positions of the antennas, leading to a higher transmission rate for data offloading~\cite{FAS, MIMO_cap_cha_for_MA}. As a result, a significant lower offloading latency can be obtained. 
Pioneering works on the FA-assisted LM-embedded wireless network have explored the benefits that FA brings to the model training and convergence performance~\cite{dl_fl2,dl_fl3}. However, these results primarily focused on the training phase for single-antenna users but did not investigate the inference phase and the more practical scenarios involving multi-antenna users.

Due to the discrete quantization model employed for latency reduction, the network design problem becomes a non-convex mixed-integer problem, which is generally difficult to solve. Worse still, the FA deployment would significantly exacerbate the non-convexity due to the strongly coupling variables in antenna distance constraint~\cite{MIMO_cap_cha_for_MA}.
The situation is even more challenging if there exists multi-antenna users. Yet, a systematic and tractable approach for handling the complicated discrete non-convex optimization in FA-assisted LM-embedded wireless network and jointly optimizing the network latency and model accuracy has not been studied before.

To overcome the above challenges and fill this gap, this paper for the first time reveals an equivalent transformation to handle the quantization model for FA-assisted LM-embedded MIMO network. To be specific, we first formulate a novel loss function, which jointly combines the network latency and the quality of quantized features. Then, an efficient optimization algorithm is proposed under the block coordinate descent (BCD) framework, where the subproblems can be solved with closed-form solutions, sequential update algorithms, and the Riemannian alternating direction method of multipliers (RADMM)~\cite{riemannian_admm}, respectively. Finally, extensive simulation results are provided to show the convergence behavior of the proposed algorithm, and the performance gains from the proposed FA-assisted LM-embedded network over the benchmark networks in terms of network latency and peak signal to noise ratio (PSNR).

\section{System Model and Problem Formulation}~\label{model}
\subsection{System Model}

\begin{figure}
    \centering    \includegraphics[width=0.65\linewidth]{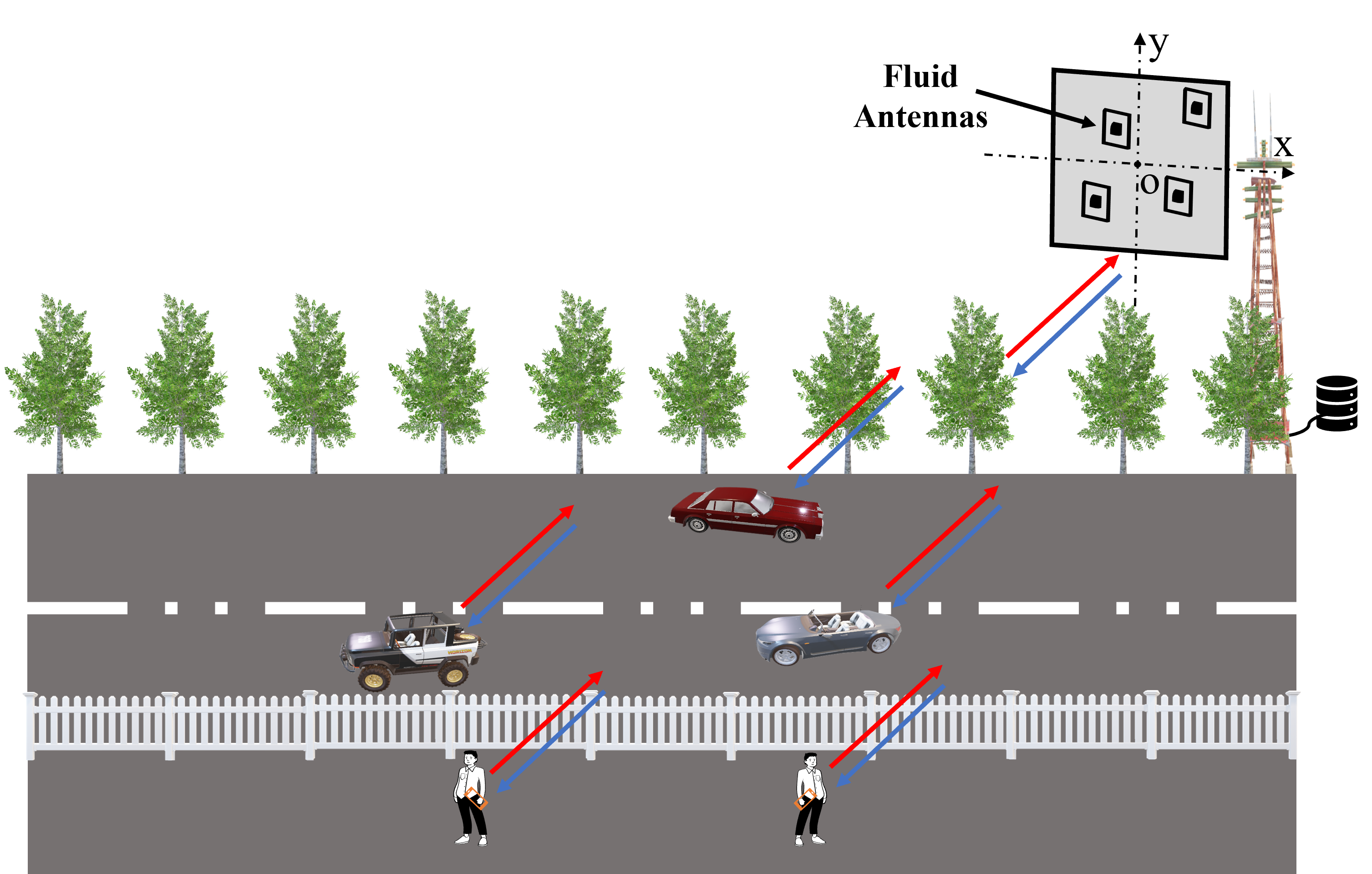}
    \caption{Illustration of the FA-assisted wireless system.}
    \label{fig: system}
\end{figure}

This paper explores an LM-embedded network serving multi-antenna users for their edge inference tasks, as shown in Fig.~\ref{fig: system}. Specifically, the network consists of a base station (BS) equipped with $M$ FAs and a server computing for LM inference. There are $U$ users, each equipped with $A$ fixed-position antennas (FPAs). For simplicity, we define $\mathcal{U} = \{1,2,\ldots,U\}$ as the set of users, $\mathcal{A} = \{1,2,\ldots,A\}$ as the set of antennas of each user, and $\mathcal{M} = \{1,2,\ldots,M\}$ as the set of FAs of the BS.
In the proposed wireless system, the FAs can move freely within their antenna panel, and the system employs space-division multiple access to enable users to communicate concurrently with the BS~\cite{MIMO_cap_cha_for_MA}. Therefore, the number of users' antennas does not exceed the number of FAs of the BS, i.e., $UA \leq M$. It is assumed that the time required for adjusting FA positions is negligible~\cite{MIMO_cap_cha_for_MA}. Let $\mathbf{t}_{u,a} \triangleq \left[x_{t_{u,a}}, y_{t_{u,a}} \right]^T$ represent the position of the $a$-th antenna of the $u$-th user, $\forall a \in \mathcal{A}$ and $\forall u \in \mathcal{U}$. These antenna positions can be aggregated to form a matrix, denoted as $\mathbf{T} = \left[ \mathbf{t}_{1,1}, \mathbf{t}_{1,2}, \ldots,\mathbf{t}_{1,A}, \mathbf{t}_{2,1}, \ldots, \mathbf{t}_{U,A} \right]$. Similarly, the position of the $m$-th FA of the BS can be denoted as $\mathbf{r}_{m} = \left[x_{r_m}, y_{r_m} \right]^T$, $\forall m \in \mathcal{M}$. The corresponding matrix is defined as $\mathbf{R} = \left[ \mathbf{r}_{1}, \mathbf{r}_{2}, \ldots, \mathbf{r}_{M} \right]$.

The narrow-band quasi-static and far-field model is used to formulate the channel, and the channel can be expressed by a function with respect to FA positions~\cite{MIMO_cap_cha_for_MA}. In particular, the array response matrix of the BS is denoted by $\mathbf{F} \left(   \mathbf{R} \right) = \left[  \mathbf{f}_1(\mathbf{R}), \mathbf{f}_2(\mathbf{R}), \dots, \mathbf{f}_U(\mathbf{R}) \right]$, and $\mathbf{f}_u(\mathbf{R})$ is given by
\begin{align} 
\label{array response vector}
\mathbf{f}_u(\mathbf{R}) = \left[ \exp\left\{j \frac{2 \pi}{\lambda} \rho_u(\mathbf{r}_1)\right\}, \ldots, \exp\left\{j \frac{2 \pi}{\lambda} \rho_u(\mathbf{r}_M)\right\}\right]^T,
\end{align}
% \begin{align} 
% \label{array response vector}
% \mathbf{f}_u(\mathbf{R}) = \left[ \exp\left\{j \frac{2 \pi}{\lambda} \rho_u(\mathbf{r}_1)\right\}, \exp\left\{j \frac{2 \pi}{\lambda} \rho_u(\mathbf{r}_2)\right\}, \right.\nonumber \\
% \left.\ldots, \exp\left\{j \frac{2 \pi}{\lambda} \rho_u(\mathbf{r}_M)\right\}\right]^T,
% \end{align}
where $\lambda$ represents the carrier wavelength, and $\rho_u(\mathbf{r}_m)\triangleq x_{r_m}\sin \theta_r^u \cos \phi_r^u + y_{r_m} \cos \theta_r^u$ with $\theta_r^u$ and $\phi_r^u$ denoting the elevation and azimuth angles of arrival of the $u$-th user. The array steering matrix of the users is given by $\mathbf{G} = \left[  \mathbf{g}_1, \mathbf{g}_2, \dots, \mathbf{g}_U  \right]$, where the $\mathbf{g}_u$ is defined as $\mathbf{g}_u = \left[ \exp\left\{j \frac{2 \pi}{\lambda} \rho_{u,1}\right\},~~\ldots,~~\exp\left\{j \frac{2 \pi}{\lambda} \rho_{u,A}\right\}\right]^T$,
and $\rho_{u,a} \triangleq x_{t_{u,a}} \sin \theta_t^u \cos \phi_t^u + y_{t_{u,a}} \cos \theta_t^u$ with $\theta_t^u$ and $\phi_t^u$ denoting the elevation and azimuth angles of departure of the $u$-th user. Based on this,~\cite{Joint_BFandAMD_forMA_statCSI,MIMO_cap_cha_for_MA} defined the channel as $\mathbf{H}\left(\mathbf{R}\right)$.

During uplink transmission, users transmit their feature embeddings to the server, and the signal at the BS can be expressed as \begin{align}
\label{uplink received signal}
    \mathbf{y}_{ul} = \mathbf{W}^H\mathbf{H}\left(\mathbf{R}\right)\mathbf{P} ^{\frac{1}{2}}\mathbf{s}+\mathbf{W}^H\mathbf{n},
\end{align}
where $\mathbf{W} = \left[\mathbf{w}_{1,1}, \mathbf{w}_{1,2}, \ldots, \mathbf{w}_{U,A}\right]$ represents the receive beamforming matrix. The channel of entire system is given by $\mathbf{H} \left( \mathbf{R} \right) = \left[ \mathbf{H}_1, \mathbf{H}_2, \dots, \mathbf{H}_U \right]$ with $\mathbf{H}_u$ being the channel of user $u$. The matrix $\mathbf{P} = \operatorname{diag} \left[ p_{1,1}, p_{1,2}, \ldots, p_{U,A} \right]$ is a diagonal matrix representing users' transmit power, with its elements $\{ p_{u,a} \}_{u=1,a=1}^{U,A}$ denoting the power allocated to the $a$-th antenna of user $u$. It is assumed that the equal power distribution is adopted by all users to their antennas. The transmitted signal $\mathbf{s}$ has unit power, $\mathbb{E}\left\{\mathbf{s} \mathbf{s}^H\right\} = \mathbf{I}_{U\times A}$, and $\mathbf{n} \sim \mathcal{CN} \left(\mathbf{0},\sigma^2\mathbf{I}_M\right)$ is the additive white Gaussian noise with variance being $\sigma^{2}$. Accordingly, the uplink transmission rate of the $u$-th user is
\begin{align}
\label{uplink rate-u}
    R_{ul,u} = \log_2 \operatorname{det} \left( \mathbf{I}_A + \mathbf{W}_u^H \mathbf{H}_u  \mathbf{P}_u^{\frac{1}{2}} \mathbf{Z}_{ul,u}^{-1} \mathbf{P}_u^{\frac{1}{2}H} \mathbf{H}_u^H \mathbf{W}_u\right),
\end{align}
where $\mathbf{W}_u = \left[\mathbf{w}_{u, 1}, \mathbf{w}_{u, 2} \ldots, \mathbf{w}_{u, A}\right]$ refers to the receive beamforming of the $u$-th user, and $\mathbf{P}_u^{\frac{1}{2}} = \operatorname{diag}\left[\sqrt{p_{u,1}}, \sqrt{p_{u,2}} \ldots, \sqrt{p_{u,A}}\right]$ is the transmit power of the $u$-th user. Furthermore, $\mathbf{Z}_{ul,u}$ denotes the covariance matrix of interference plus noise of user $u$, which is derived as $\mathbf{Z}_{ul,u} = \sigma^2 \mathbf{W}_u^H \mathbf{W}_u + \sum_{k \in \mathcal{U}, k \neq u} \mathbf{W}_u^H \mathbf{H}_k \mathbf{P}_k \mathbf{H}_k^H \mathbf{W}_u$.

\subsection{The Model of LM-based Inference} 

\begin{figure}
    \centering
    \includegraphics[width=0.9\linewidth]{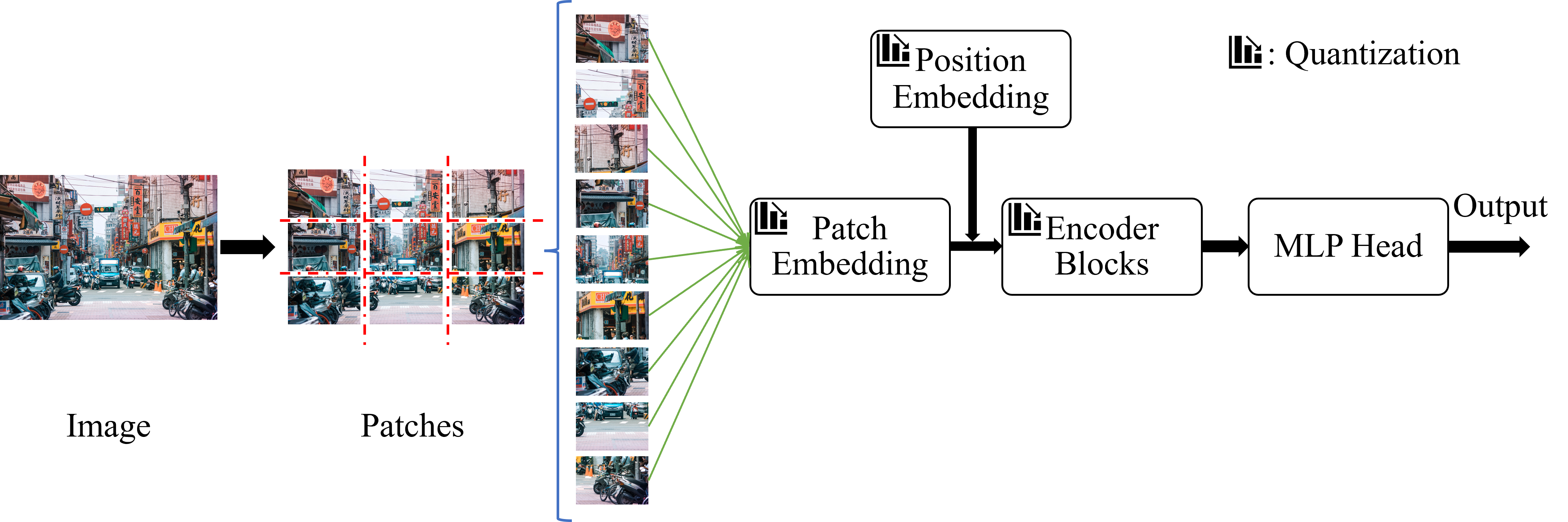}
    \caption{Illustration of the LM quantization.}
    \label{fig: quantization_of_LM}
\end{figure}

It is assumed that all users are executing their inference tasks using the fixed model. Generally, the data collected by users must be processed by the embedding layer before being fed into the main body of LM. The size of embedding layer is significantly smaller than that of the subsequent LM layers, making it feasible to deploy this layer on the client side. Based on this knowledge, the quantization in this paper aims at the embedding layer or its subsequent layers, as illustrated in Fig.~\ref{fig: quantization_of_LM}.

The overall latency of the proposed system is given by $T = \sum_{u=1}^U T_{fo,u} + T_{si,u}$, where $T_{fo,u}$ and $T_{si,u}$ are respectively the latency of feature offloading and server inference.\footnote{Once the server finishes the computation for the received data, the inference results will be distributed to the users. It is assumed that the BS is powered by the grid, allowing for the transmission of the downlink signal with relatively high power, so the downlink latency is not considered.} Once the embeddings are obtained, user $u$ will offload them to the server. Assuming that all users offload their data within the same time slot, the feature offloading latency is given by
\begin{align}
    T_{fo,u} = \frac{q_u D_u}{b R_{ul,u}},
\end{align}
where $q_u \in \{4,8,16,32\}$ represents the number of quantization digits used to encode the $u$-th user's features~\cite{LM-quantize}, and $\mathbf{q}$ denotes the collection of all $q_u$ values. $D_u$ refers to the size of the offloaded features, and $b$ represents the transmission bandwidth. Upon receiving the data from user $u$, the server fulfills the subsequent inference, and this latency is calculated as~\cite{UA_RA_LLM_MEC_6G}
\begin{align}
    T_{si,u} = \frac{c_s d_u}{f_{s,u}},
\end{align}
where $d_u$ characterizes the LM for the $u$-th user by its parameter size, $c_s$ signifies the number of processing unit (PU) cycles required to process the data for one parameter at the server, and $f_{s,u}$ represents the PU frequency of the server allocated to user $u$ with $\mathbf{f}_{s}$ being the corresponding vector.

Considering uniform quantization~\cite{LM-quantize}, the quantization step size and the quantization digits satisfy $\Delta = \frac{ x_{\text{max}} - x_{\text{min}} }{2^{q_u}-1}$,
where $\Delta$ is the step size of quantization, $x_{\text{min}}$ and $x_{\text{max}}$ denote the minimal and maximal values of the data. It is assumed that the quantization error follows a uniform distribution over $\left[ -\Delta/2, \Delta/2 \right]$, so the power of quantization error is given by $\sigma_e^2=\Delta^2/12$. The peak signal-to-noise ratio (PSNR) is a crucial metric for assessing the performance of LM, and revealing the positive correlation between LM performance and feature quality~\cite{FL-weibin1}. The PSNR of the $u$-th user is given by
\begin{align}
    PSNR_u \triangleq 10\operatorname{log}_{10} \left( \frac{\sigma_{x_{\text{max}}}^2}{\sigma_e^2}  \right),
\end{align}
where $\sigma_{x_{\text{max}}}$ refers to the maximal value of features, and the total PSNR of the system is $PSNR = \sum_{u=1}^U PSNR_u$.

\subsection{Problem Formulation}   
% 我在后面仿真setting的介绍里把天线面板的边长定义成了L，为了区分，我在这里把目标函数定义成了Loss()而不是L()

To meet the user’s comprehensive requirements for network latency and inference accuracy, our objective aims at minimizing the loss function which is the combination of the total network latency and model accuracy. Mathematically, the loss function is defined as
\begin{align}
    \operatorname{Loss}(\{\mathbf{r}_m\}_{m=1}^M, \{\mathbf{w}_{u,a}\}_{u=1,a=1}^{U,A}, \mathbf{f}_{s}, \mathbf{q}) = T-\alpha\cdot PSNR,
\end{align}
 where the positive factor $\alpha$ refers to the sensitivity of LM performance to quantization~\cite{HAWQ}. Accordingly, the optimization problem is formulated as
\begin{align}
\mathcal{P}:&\min_{ \{\mathbf{r}_m\}_{m=1}^M, \{\mathbf{w}_{u,a}\}_{u=1,a=1}^{U,A}, \mathbf{f}_{s}, \mathbf{q}} \quad \operatorname{Loss} \nonumber \\ 
&~~~~~\text { s.t. }
\sum_{u=1}^U f_{s,u} \leq f_{\text{ser}}, \label{cpu frequency}\\
& ~~~~~~~~~~~ \mathbf{r}_m \in \mathcal{C}_{\text{R}}, ~\forall m \in \mathcal{M}, \label{BS given region}\\
& ~~~~~~~~~~ \left\|\mathbf{r}_m-\mathbf{r}_l\right\|_2 \geq D, ~\forall m,l \in \mathcal{M}, ~ m \neq l, \label{BS antenna distance} \\
& ~~~~~~~~~~~ q_u \in \{4,8,16,32\},~\forall u \in \mathcal{U}, \label{quantization digits} \\
& ~~~~~~~~~~~ \| \mathbf{w}_{u,a} \|_2 = 1, ~\forall u \in \mathcal{U},~\forall a \in \mathcal{A}. \label{receive beamforming power}
\end{align}
The constraints of this problem are detailed as follows:~\eqref{cpu frequency} ensures that the allocated computational frequency of the server does not exceed its budget $f_{\text{ser}}$. Constraint~\eqref{BS given region} restricts the movement of FAs, implying that the adjusted FA must stay within the antenna panels. Furthermore,~\eqref{BS antenna distance} enforces constraints on antenna distances to mitigate coupling effects~\cite{MIMO_cap_cha_for_MA}. Then,~\eqref{quantization digits} imposes constraints on the selection of quantization digits. Finally,~\eqref{receive beamforming power} ensures the beamforming power allocated to each user~\cite{why_manifold_beamforming}. This problem is challenging to solve because it is non-convex, mixed-integer, and variable-coupled.

\section{Solution to the Joint Optimization Problem} \label{algorithm}

It is noticed that the constraints in $\mathcal{P}$ for all variables are separable. Therefore, the BCD algorithm can be adopted
to sequentially update each variable.

\subsection{Updating $\mathbf{f}_{s}$}

When other variables are fixed, the subproblem of $\mathcal{P}$ for updating PU frequency $\mathbf{f}_{s}$ is
\begin{align}
    \mathcal{P}1:&~~\min_{\mathbf{f}_{s}} \quad f_1 \left( \mathbf{f}_{s} \right) \triangleq \sum_{u=1}^U \frac{c_s d_u}{f_{s,u}} \nonumber\\ 
    &~~\text { s.t. } \quad \eqref{cpu frequency}.  \nonumber
\end{align}
% By calculating the second-order derivatives of $f_1 \left( \mathbf{f}_{s} \right)$ regarding to all elements of $\mathbf{f}_{s}$, we obtain
% \begin{align}
%     \frac{\partial^2 f_1 \left( \mathbf{f}_{s} \right)}{\partial f_{s,u}^2} &= \frac{2 c_s d_u}{f_{s,u}^3} > 0, \quad \forall u \in \mathcal{U}, \\
%     \frac{\partial^2 f_1 \left( \mathbf{f}_{s} \right)}{\partial f_{s,u} \partial f_{s,k}} &= 0, \quad \forall u,k \in \mathcal{U}, ~~u \neq k.
% \end{align}
Since the Hessian matrix of $f_1 \left( \mathbf{f}_{s} \right)$ is positive definite, and the feasible set determined by~\eqref{cpu frequency} is convex, $\mathcal{P}1$ is convex. Then, a closed-form solution of $f_{s,u}$ can be derived based on Karush-Kuhn-Tucker (KKT) condition and is given by the
following lemma, which is proved in Appendix~\ref{appendix: optimal f_s_u}.

\begin{lemma} \label{lemma: optimal f_s_u}
The optimal solution of $f_{s,u}$ is
\begin{align} 
    \label{optimal_f}
    f_{s,u}^{\star} = \frac{\sqrt{c_s d_u} \cdot f_{\text{ser}}}{\sum_{k=1}^U \sqrt{c_s d_k}},~\forall k \in \mathcal{U}.
\end{align}
\end{lemma}

\subsection{Updating $\mathbf{q}$}

When other variables are fixed, the subproblem of $\mathcal{P}$ for updating quantization digits $ \mathbf{q}$ is
\begin{align}
    \mathcal{P}2:&~~\min_{\mathbf{q}} \quad f_2 \left( \mathbf{q} \right) \triangleq \sum_{u=1}^U \left( \frac{q_u D_u}{b R_{ul,u}} \right. \nonumber \\ 
    &~~~~~~~~~~~~~\left. -\alpha \cdot 10 \operatorname{log}_{10} \left[ \frac{12 \sigma_{x_{\text{max}}}^2 \cdot \left(2^{q_u}-1\right)^2}{\left( x_{\text{max}} - x_{\text{min}} \right)^2}  \right] \right) \nonumber \\ 
    &~~\text { s.t. } \quad \eqref{quantization digits}.    \nonumber
\end{align}
To handle this subproblem, we provide Lemma~\ref{lemma: optimal q_u}, and the proof of this lemma can be checked in Appendix~\ref{appendix: proof q}.

\begin{lemma} \label{lemma: optimal q_u}
    The function $f_2 \left( \mathbf{q} \right)$ is strictly convex when $q_u \in \left(0, +\infty \right),\forall u \in \mathcal{U}$, and the minimal value of $f_2 \left( \mathbf{q} \right)$ on $\left(0, +\infty \right)$ corresponds to the $q_u$ given by 
    \begin{align}
        q_u = \left\{ 
        \begin{aligned}
        &\log_2 \frac{D_u\ln{10}}{z_u},~&z_u > 0, \\
        &~~~~~~~~+\infty,~&z_u \leq 0, 
        \end{aligned}
        \right.
    \end{align}
    where $z_u = D_u\ln{10} - 20\alpha \ln{2} \cdot b R_{ul,u}$.
\end{lemma}

Then, we project the obtained $q_u$ to the feasible set of $\mathcal{P}2$ by solving the following optimization problem:
\begin{align} \label{qu_projection}
   q_u^{\star} = \operatorname{argmin}_{x \in \left[ 4,8,16,32 \right]} |x - q_u |. 
\end{align}

\subsection{Updating $\{\mathbf{w}_{u,a}\}_{u=1,a=1}^{U,A}$}

When other  variables are fixed, the subproblem of $\mathcal{P}$ for updating receive beamforming $\{\mathbf{w}_{u,a}\}_{u=1,a=1}^{U,A}$ is
\begin{align}
    \mathcal{P}3:&~~\min_{\{\mathbf{w}_{u,a}\}_{u=1,a=1}^{U,A}} \quad f_3 \left( \{\mathbf{w}_{u,a}\}_{u=1,a=1}^{U,A} \right) \triangleq \sum_{u=1}^U \frac{q_u D_u}{b R_{ul,u}}  \nonumber \\ 
    &~~~~~~~~\text { s.t. } \quad \eqref{receive beamforming power}.    \nonumber
\end{align}
Due to the non-convex oblique manifold constraint~\eqref{receive beamforming power}, $\mathcal{P}3$ is a nontrivial problem. To handle it, this paper resorts to the RADMM. The above $\mathcal{P}3$ can be reformulated as~\cite{riemannian_admm}
\begin{align}
    \mathcal{P}3-(\text{RADMM}):&~~\min_{\mathbf{W}, \mathbf{W}^{\prime}} \quad f_3 \left( \mathbf{W} \right) + I_{\mathcal{S}} \left( \mathbf{W}^{\prime} \right) \nonumber \\ 
    &~~~~\text { s.t. } \quad~ \mathbf{W} = \mathbf{W}^{\prime},
\end{align}
where $\mathbf{W}^{\prime}$ and $I_{\mathcal{S}} \left( \mathbf{W}^{\prime} \right)$ are respectively the auxiliary variable  and indicator function. In particular, $I_{\mathcal{S}} \left( \mathbf{W}^{\prime} \right)$ is expressed as
\begin{align}
    I_{\mathcal{S}} \left( \mathbf{W}^{\prime} \right) = \left\{ 
    \begin{aligned}
    & ~~0, ~ \mathbf{W}^{\prime} \in \mathcal{S}, \\
    & +\infty, ~  others, 
    \end{aligned}
    \right.
\end{align}
where $\mathcal{S} \triangleq \{   \mathbf{W}^{\prime}: \| \mathbf{w}_{u,a}^{\prime} \|_2 = 1, \forall u \in \mathcal{U}, \forall a \in \mathcal{A} \}$. The $(k+1)$-th update of $\mathcal{P}3-(\text{RADMM})$ is accomplished by the steps shown in \eqref{update_w} - \eqref{update_lambda}, where $\circ$ refers to the Hadamard product, and $\rho$ is the penalty factor.
\begin{figure*}
\begin{align}
    \mathbf{W}^{k+1} &= \operatorname{argmin}_{\mathbf{W}} \quad  \sum_{u=1}^U \frac{q_u D_u}{b R_{ul,u}} + \mathbf{1}_M^T \left[ \boldsymbol{\lambda}^k \circ \left(  \mathbf{W} - \mathbf{W}^{\prime k} \right) \right] \mathbf{1}_N + \frac{\rho}{2} \| \mathbf{W} - \mathbf{W}^{\prime k} \|_F^2, \label{update_w} \\
    \mathbf{W}^{\prime k+1} &= \operatorname{argmin}_{\mathbf{W}^{\prime} \in \mathcal{S}} \quad g \left( \mathbf{W}^{\prime} \right) \triangleq \mathbf{1}_M^T \left[ \boldsymbol{\lambda}^k \circ \left(  \mathbf{W}^{k+1} - \mathbf{W}^{\prime} \right) \right] \mathbf{1}_N + \frac{\rho}{2} \| \mathbf{W}^{k+1} - \mathbf{W}^{\prime} \|_F^2, \label{update_w'} \\
    \boldsymbol{\lambda}^{k+1} &= \boldsymbol{\lambda}^k + \rho \left(  \mathbf{W}^{k+1} - \mathbf{W}^{\prime k+1}\right). \label{update_lambda}
\end{align}
\hrulefill
\end{figure*}
Obviously, the update of $\mathbf{W}$ can be handled by gradient-based methods, and the update of $\boldsymbol{\lambda}$ is quite simple. The update of $\mathbf{W}^{\prime}$ on the intricate manifold will be accomplished by the Riemannian gradient method. Specifically, taking the $(k+1)$-th iteration as an example, the direction of updating $\mathbf{W}^{\prime}$ is given by $\mathbf{v}^{k+1} = \nabla_{\mathbf{W}^{\prime}} g \left( \mathbf{W}^{\prime} \right) - \Re \{ \nabla_{\mathbf{W}^{\prime}} g \left( \mathbf{W}^{\prime} \right) \circ \mathbf{W}^{\prime k \ast} \} \circ \mathbf{W}^{\prime k}$~\cite{AM_HP_HW_MIMO},
where $\nabla_{\mathbf{W}^{\prime}} g \left( \mathbf{W}^{\prime} \right)$ is the Euclidean gradient with respect to $\mathbf{W}^{\prime}$. $\mathbf{W}^{\prime}$ is then updated by a retraction shown as $\mathbf{W}^{\prime k+1} = \frac{\mathbf{W}^{\prime k} - \gamma \mathbf{v}^{k+1}}{\| \mathbf{W}^{\prime k} - \gamma \mathbf{v}^{k+1} \|}$, where $\gamma$ is the step size.

\subsection{Updating $\{\mathbf{r}_m\}_{m=1}^M$} 

When other variables are fixed, the subproblem of $\mathcal{P}$ for updating the FA positions $\{\mathbf{r}_m\}_{m=1}^M$ is
\begin{align}
    \mathcal{P}4:&~~\min_{\{\mathbf{r}_m\}_{m=1}^M} \quad f_4 \left( \{\mathbf{r}_m\}_{m=1}^M \right) \triangleq \sum_{u=1}^U \frac{q_u D_u}{b R_{ul,u}} \nonumber \\ 
    &~~\text { s.t. } \quad \eqref{BS given region} \text{ and } \eqref{BS antenna distance}.    \nonumber
\end{align}
To tackle the non-convex and coupled constraints~\eqref{BS antenna distance}, we decouple the $\mathcal{P}4$ by sequentially updating the position of the $m$-th FA in $\mathcal{P}4-\text{m}$.
\begin{align}
    \mathcal{P}4-\text{m}:&~~\min_{\mathbf{r}_m} \quad f_4 \left( \{\mathbf{r}_m\}_{m=1}^M \right) \nonumber \\ 
    &~~\text { s.t. } ~~ \mathbf{r}_m \in \mathcal{C}_{\text{R}}, \label{BS given region-m}\\
    & ~~~~~~~~ \left\|\mathbf{r}_m-\mathbf{r}_l\right\|_2 \geq D, ~~\forall l \in \mathcal{M}, ~m \neq l. \label{BS antenna distance-m}
\end{align}
$\mathcal{P}4-\text{m}$ can be efficiently handled by any sequential updating methods~\cite{MIMO_cap_cha_for_MA, GD1}. We here use the method proposed in~\cite{MIMO_cap_cha_for_MA} to handle $\mathcal{P}4$.

\subsection{The Overall Algorithm and Discussions}

The overall algorithm for tackling $\mathcal{P}$ is summarized in Algorithm~\ref{a1}. The algorithm solves all subproblems iteratively, until the relative change of the objective function of $\mathcal{P}$ below a predefined convergence threshold $\epsilon$.
\begin{algorithm}[t] 
\caption{The Overall Algorithm for Solving $\mathcal{P}$ } 
\begin{algorithmic}[1] \label{a1}
\STATE Initialize all variables
\REPEAT

\STATE \textbf{for} $u=1,\ldots,U$ \textbf{do}
\STATE ~~~~Update $f_{s,u}$ based on~\eqref{optimal_f}.
\STATE \textbf{end for}

\STATE Update $\mathbf{q}$ by the Lemma~\ref{lemma: optimal q_u} and~\eqref{qu_projection}.

\REPEAT
\STATE ~~Update $\{\mathbf{w}_{u,a}\}_{u=1,a=1}^{U,A}$ via sequentially update $\mathbf{W}$, \\~~$\mathbf{W}^{\prime}$, and $\boldsymbol{\lambda}$ based on~\eqref{update_w},~\eqref{update_w'}~\text{and}~\eqref{update_lambda}.
\UNTIL Stopping criterion is satisfied.

\STATE Update $\{\mathbf{r}_m\}_{m=1}^M$ by sequentially solving $\mathcal{P}4-\text{m}$.

\UNTIL Stopping criterion is satisfied.	
\end{algorithmic}
\end{algorithm}
The convergence of the proposed Algorithm~\ref{a1} is analyzed as follows. The BCD of all variables guarantees that the Algorithm~\ref{a1} produces a non-increasing sequence of objective values of $\mathcal{P}$ during the iterations. Since for any fixed $\alpha < \infty$, the objective function of $\mathcal{P}$ has finite lower bound, the objective value is guaranteed to converge. Moreover, since the solutions of all blocks satisfy the first-order optimal condition, the proposed algorithm is guaranteed to converge to a stationary point at least. The complexity of Algorithm~\ref{a1} is also provided here. The complexity for solving $\mathcal{P}1$ and $\mathcal{P}2$ are both $\mathcal{O}(U)$. Then, the complexities for solving $\mathcal{P}3$ and $\mathcal{P}4$ are given in~\cite{riemannian_admm} and~\cite{MIMO_cap_cha_for_MA}, respectively. Overall, the complexity of Algorithm~\ref{a1} is given by the sum of all above terms.

\section{Numerical Results} \label{simulation}

In this section, we provide extensive results to demonstrate the performance of the proposed algorithm. The stopping criteria of the Algorithm~\ref{a1} are set as the relative change of the respective objective function values being smaller than $10^{-3}$. The wireless system consists of $U=3$ users, and we consider the user and BS are equipped with $A = 2$ and $M = 6$ antennas, respectively. The FAs' antenna penal $\mathcal{C}_{\text{R}}$ is set as an $L\times L$ square area. The minimum distance between FAs for avoiding the coupling effect is set as $D = \lambda/2$, and $\lambda=0.05$. The channel of user $u$ is given by $\mathbf{H}_u\left(\mathbf{R}\right) = \sqrt{\frac{K_u}{K_u+1}} \overline{\mathbf{H}}_u\left(\mathbf{R}\right) + \sqrt{\frac{1}{K_u+1}} \widetilde{\mathbf{H}}_u\left(\mathbf{R}\right)$, where $K_u = 1$ is the Rician factor. The line-of-sight channel $\overline{\mathbf{H}}_u$ is given by $\overline{\mathbf{H}}_u\left( \mathbf{R} \right) = \beta_u \mathbf{f}_u\left(\mathbf{R}\right) \mathbf{g}_u^H$, where $\beta_u$ is the complex path gain, and the non-line-of-sight channel $\widetilde{\mathbf{H}}_u$ with $N=5$ scatterers is given by $\widetilde{\mathbf{H}}_u \left(\mathbf{R}\right) = \sum_{n=1}^N \gamma_n \mathbf{f}_n\left(\mathbf{R}\right) \mathbf{g}_n^H$, where $\gamma_n \sim \mathcal{CN}\left( 0,1 \right)$ is the spatial spreading factor~\cite{Joint_BFandAMD_forMA_statCSI}. The elevation and azimuth angles are assumed to be i.i.d. variables uniformly distributed over $\left[0, \pi \right)$. The SNR of transmission is set as $P/\sigma^2 = 10dB$, and $\sigma^2=1$. All the results are obtained by Monte-Carlo simulations on Matlab-R2024a. The penalty factor $\rho$ is initialized as $5$ and increased by a factor of $1.2$ for $\mathcal{P}3$, and the sensitivity factor $\alpha$ is set as $0.02$ for an LM with $1$ billion parameters.

% \begin{figure*}
% 	\centering
%  \subfigure[]{\label{111} 
% 		\includegraphics[width=1.5in]{sim/circle1.png}} 
%    \subfigure[]{\label{222} 
% 		\includegraphics[width=1.5in]{sim/circle2.png}} 
% 	\subfigure[]{\label{333} 
% 		\includegraphics[width=1.5in]{sim/circle3.png}}  
% 	\subfigure[ ]{
% 		\label{666} 
% 		\includegraphics[width=1.5in]{sim/circle4.png}}
%   	\caption{$m=1$ and $M=4$. The red lines denote the minimal distance between $\mathbf{r}_1$ and $\mathbf{z}_1$. (a) Illustration of case 1; (b) and (c) Illustration of case 2; (d) Illustration of case 3.} 
% \end{figure*}

\begin{figure}
    \centering
    \includegraphics[width=0.65\linewidth]{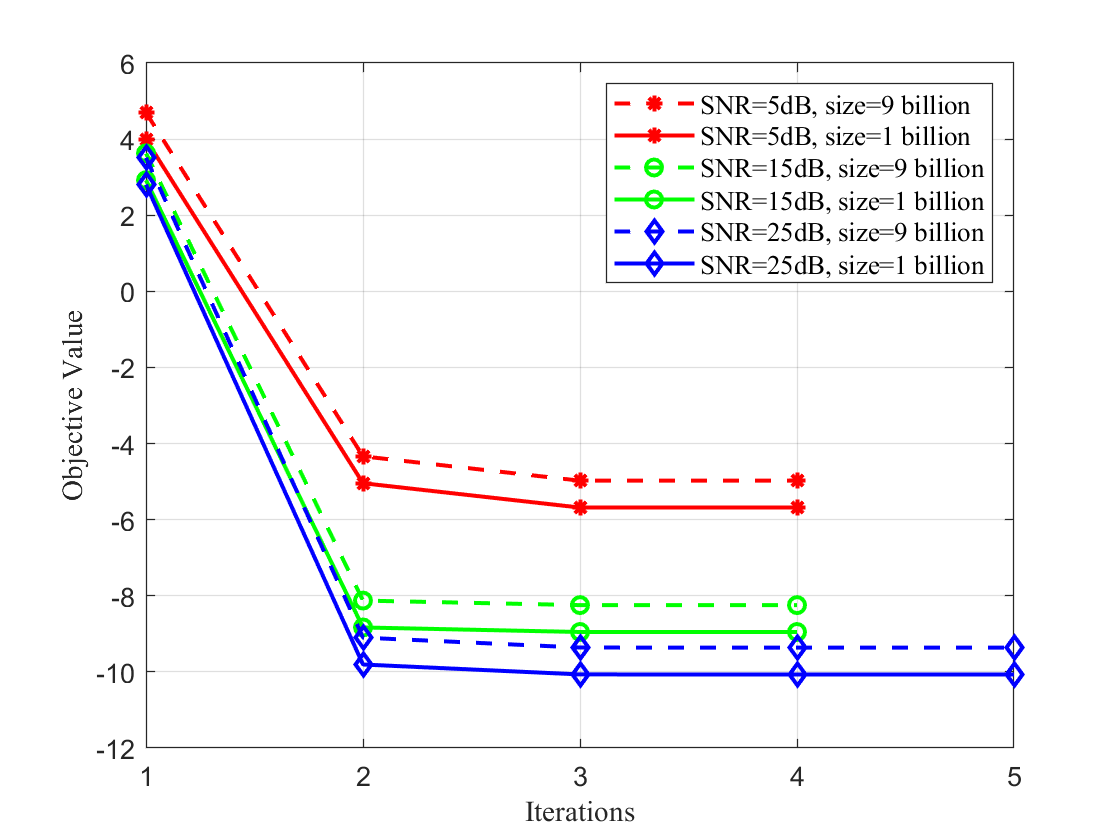}
    \caption{The convergence behavior of the objective function of $\mathcal{P}$.}
    \label{fig: convergence}
\end{figure}

Firstly, we demonstrate the convergence of the proposed algorithm with $L = 5\lambda$. As shown in Fig.~\ref{fig: convergence}, the simulation results under different SNRs and model sizes show that the objective value of $\mathcal{P}$ monotonically decreases and converges to stable values within $5$ iterations, which confirms the efficacy of the proposed algorithm. 
Additionally, the computational intensity increases with the growing number of LM parameters, leading to a higher converged objective function value.

Since the various tasks of different users often have distinct accuracy requirements, we present the relative changes in latency and PSNR by adjusting the sensitivity parameter of LM performance to quantization $\alpha$, revealing the relationship between latency and inference accuracy.
\begin{figure}
    \centering
    \includegraphics[width=0.65\linewidth]{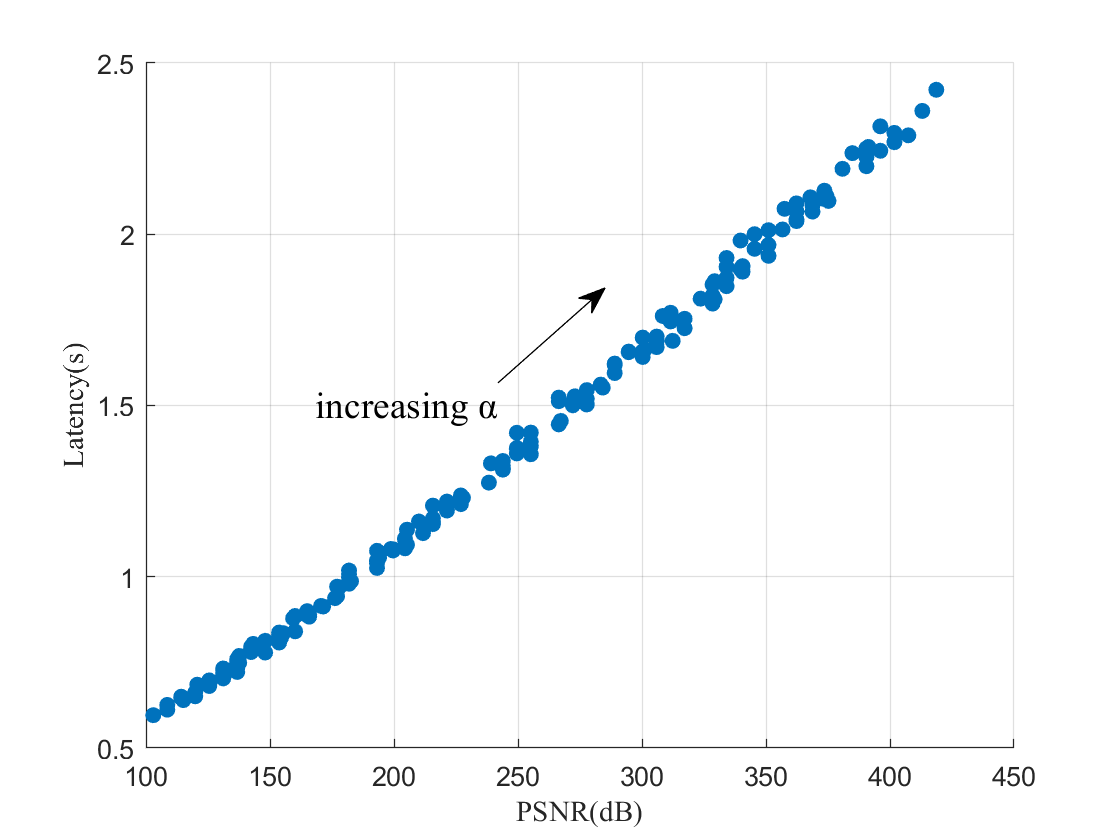}
    \caption{The illustration of PSNR and latency with increasing $\alpha$.}
    \label{fig: qVSl}
\end{figure}
As shown in Fig.~\ref{fig: qVSl}, as $\alpha$ increases, the model is more and more sensitive to the feature quality, the proposed system must use the features with higher quality for inference, which sacrifices the latency. This positive correlation between latency and PSNR indicates that the formulated loss function can characterize the proposed system.

To demonstrate the proposed system in practical, we provide the experimental results regarding a ViT-B-16-embedded wireless system. Particularly, the ViT-B-16 is an image classification network consisting of 86 million parameters~\cite{vit}.
\begin{figure}
	\centering
    \subfigure[]{\label{vit-l} 
		\includegraphics[width=0.48\linewidth]{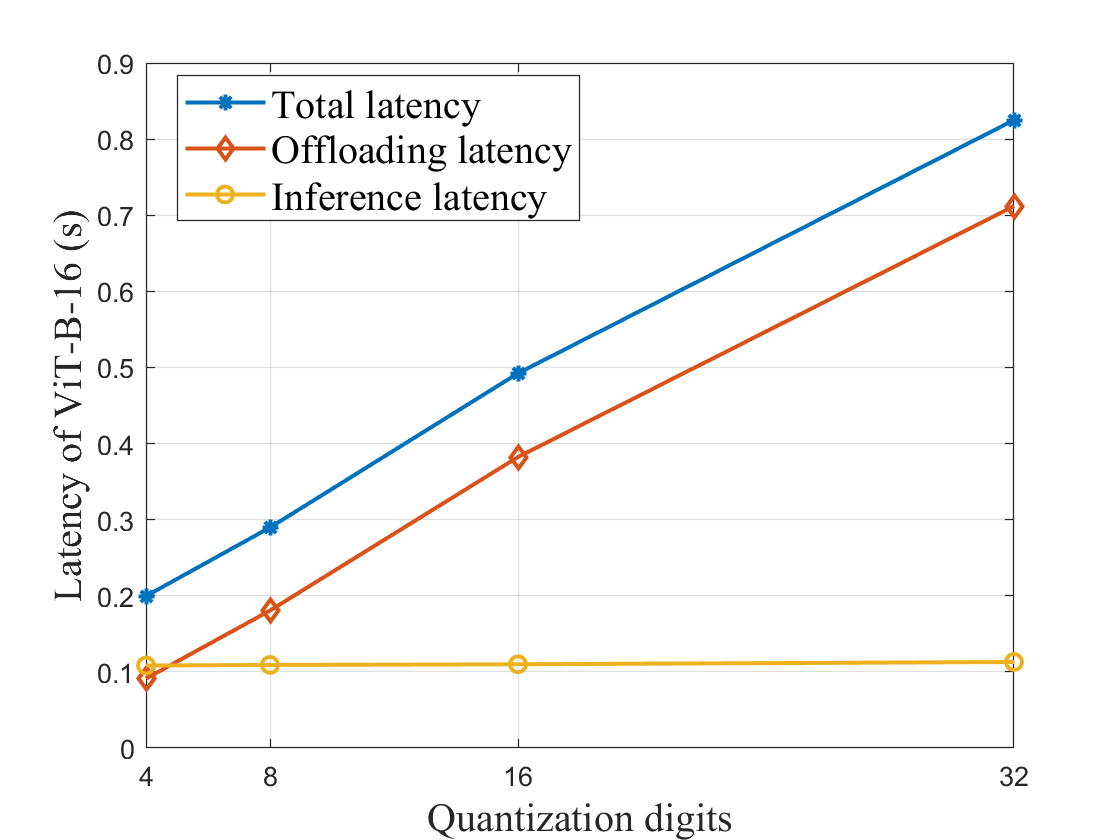}} 
    \subfigure[]{\label{vit-mae} 
		\includegraphics[width=0.48\linewidth]{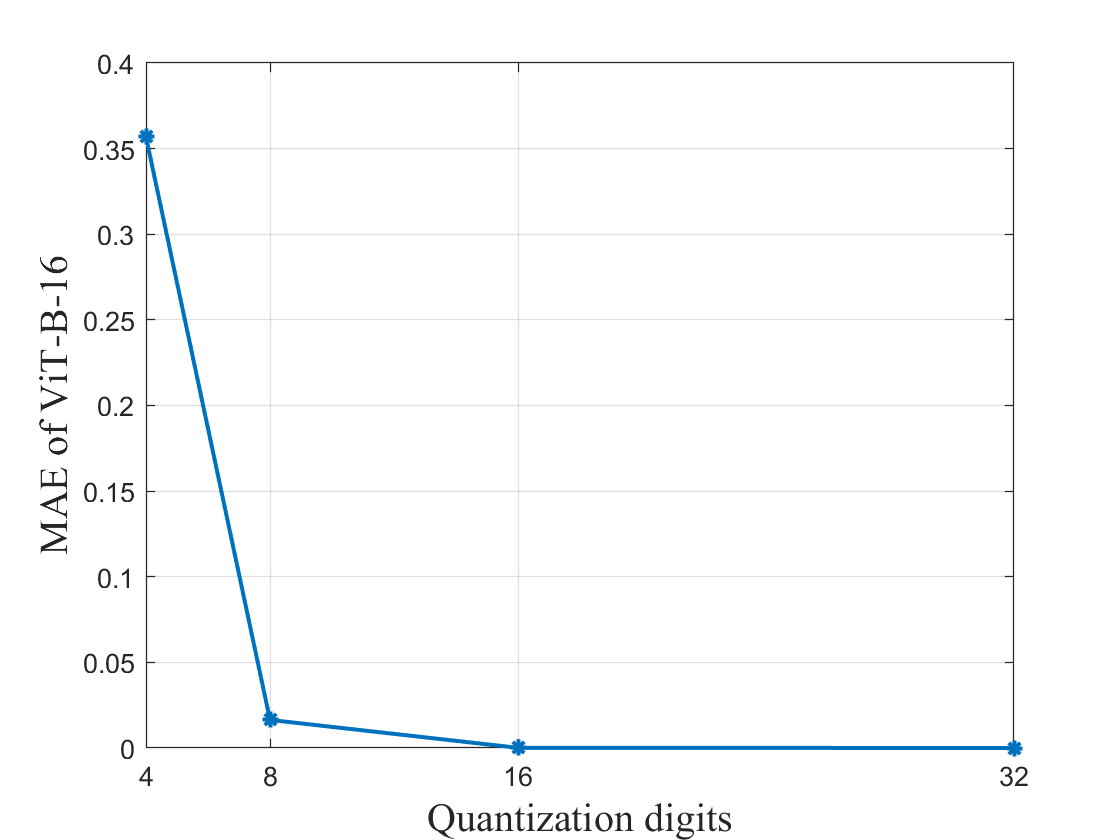}} 
  	\caption{The illustration of the performance of ViT-B-16-embedded system versus the quantization digits.}  
    \label{fig: vit}
\end{figure}
In Fig.~\ref{fig: vit}, with the increase of quantization digits, the inference mean average error (MAE) decreases rapidly, but the latency also increases significantly. In addition, the 16-bit quantized data can offer an excellent inference performance, using a larger quantization digits brings almost no extra gains, but the latency performance will be destroyed. Therefore, the proposed trade-off between latency and inference accuracy has practical significance.

% changeM.eps有图层问题，坐标轴死活放不到最下面，png就没问题
Finally, we demonstrate the performance of the proposed system versus the numbers of antennas with $U = 2$ and $L=10\lambda$. For the baselines of the FPA system, the antennas are arranged equidistantly along the diagonal of antenna panel.
\begin{figure}
    \centering
    \includegraphics[width=0.65\linewidth]{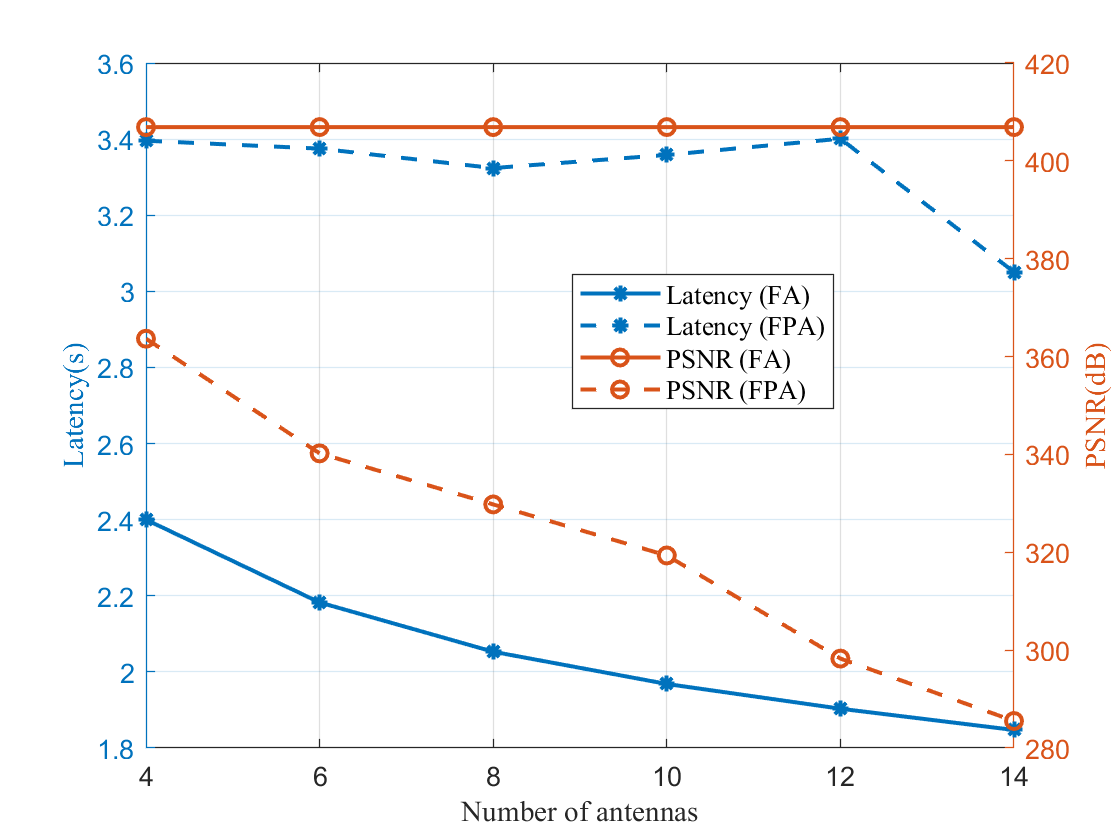}
    \caption{The illustration of the system performance versus $M$.}
    \label{fig: change M}
\end{figure}
As shown in Fig.~\ref{fig: change M}, increasing the number of FAs within a certain range can yield significant gains. However, continuously increasing the number of FAs to improve system performance eventually becomes ineffective. Meanwhile, we observe that the gain of increasing the number of antennas in the FPA system is also influenced by the arrangement of antennas, so it does not guarantee the benefit continuously. Even if the FPA system may achieve better latency performance by increasing the number of antennas, the improvement is somewhat at the expense of feature quality and inference accuracy. In contrast, systems that employ FAs can effectively address this issue, offering better performance in terms of latency and PSNR.

\section{Conclusion} \label{conclusion}

In this paper, an FA-assisted LM-embedded MIMO system has been investigated, and its performance has also been investigated. Based on the model shown in this paper, we have revealed that the trade-off between the latency and LM inference accuracy can be represented by a non-convex, variable-coupled, and mixed-integer optimization problem, and we subsequently proposed an efficient algorithm for solving this problem. This paper also provided sufficient experimental results which confirmed the excellent performance of the proposed system and algorithm.

\appendices

\section{Proof of the Lemma~\ref{lemma: optimal f_s_u}} \label{appendix: optimal f_s_u}

Based on the Lagrangian function $\mathcal{L} \left( \mathbf{f}_{s}, \mu \right) = \sum_{u=1}^U \frac{c_s d_u}{f_{s,u}} \nonumber + \mu \left( \sum_{u=1}^U f_{s,u} - f_{\text{ser}} \right)$ with $\mu$ being the Lagrangian multiplier, the KKT condition is represented by $\nabla_{f_{s,u}} \mathcal{L} \left( \mathbf{f}_{s}, \mu \right) = \frac{- c_s d_u}{f_{s,u}^2} + \mu  =0,~ \sum_{u=1}^U f_{s,u} - f_{\text{ser}} \leq 0,~ \mu \geq 0, \text{ and } \mu \left( \sum_{u=1}^U f_{s,u} - f_{\text{ser}} \right) = 0$.
% \begin{align}
%     &\nabla_{f_{s,u}} \mathcal{L} \left( \mathbf{f}_{s,u}, \mu \right) = \frac{- c_s d_u}{f_{s,u}^2} + \mu  =0,  \label{derivative of Lagrangian} \\
%     &\sum_{u=1}^U f_{s,u} - f_{\text{ser}} \leq 0, \nonumber \\
%     &\mu \geq 0, \text{ and } \mu \left( \sum_{u=1}^U f_{s,u} - f_{\text{ser}} \right) = 0. \label{mu and frequency}
% \end{align}

By examining the condition of first-order derivative, the optimal solution of $f_{s,u}$ is
\begin{align}
    f_{s,u}^{\star} = \sqrt{\frac{c_s d_u}{\mu^{\star}}}. \label{optimal f_{s,u}}
\end{align}
According to the KKT condition, $\sum_{u=1}^U f_{s,u}^{\star} - f_{\text{ser}} = 0$ holds for any $\mu > 0$. Plugging~\eqref{optimal f_{s,u}} into this equation, the closed form solution of $\mu^{\star}$ is obtained as $\mu^{\star} = \left( \frac{\sum_{u=1}^U \sqrt{c_s d_u}}{f_{\text{ser}}} \right)^2$.
% \begin{align}
%     \mu^{\star} = \left( \frac{\sum_{u=1}^U \sqrt{c_s d_u}}{f_{\text{ser}}} \right)^2. \label{optimal mu}
% \end{align}
Plugging the $\mu^{\star}$ into~\eqref{optimal f_{s,u}}, we arrive at the final result.

\section{Proof of the Lemma~\ref{lemma: optimal q_u}}   \label{appendix: proof q}

The first-order derivatives of $f_2 \left( \mathbf{q} \right)$ regarding to all elements of $\mathbf{q}$ can be obtained by
\begin{align} \label{qu_gradient}
    \frac{\partial f_2 \left( \mathbf{q} \right)}{\partial q_u} = \frac{D_u}{b R_{ul,u}} - \frac{20\alpha \cdot 2^{q_u} \ln{2}}{\left( 2^{q_u}-1 \right) \ln{10}},~\forall u \in \mathcal{U},
\end{align}
the second-order derivatives are
\begin{align}
    \frac{\partial^2 f_2 \left( \mathbf{q} \right)}{\partial q_u^2} &= \frac{20\alpha \ln{2}}{\ln{10}} \cdot \frac{2^{q_u} \ln{2}}{2^{q_u}-1} \cdot \left(\frac{1}{2^{q_u}-1}\right) >0, ~\forall u \in \mathcal{U}, \label{qu_2gradient}\\
    \frac{\partial^2 f_2 \left( \mathbf{q} \right)}{\partial q_u q_k} &= 0, ~ \forall u,k \in \mathcal{U}, ~~u \neq k. \nonumber
\end{align}
The Hessian matrix is positive definite when $q_u \in (0,+\infty)$, so the $f_2 \left( \mathbf{q} \right)$ is strictly convex on this domain.

Let~\eqref{qu_gradient} equal to $0$, we arrive at $q_u = \log_2 \frac{D_u\ln{10}}{D_u\ln{10} - 20\alpha \ln{2} \cdot b R_{ul,u}}$. The following discussion is based on whether this point exists or not. On one hand, if $D_u\ln{10} - 20\alpha \ln{2} \cdot b R_{ul,u} > 0$, the zero point of $\frac{\partial f_2 \left( \mathbf{q} \right)}{\partial q_u}$ exists. Since~\eqref{qu_2gradient} tells that the first-order derivative is monotonically increasing on $(0,+\infty)$, $q_u$ is the minimum point of $f_2 \left( \mathbf{q} \right)$. Otherwise, if $D_u\ln{10} - 20\alpha \ln{2} \cdot b R_{ul,u} \leq 0$, the zero point of $\frac{\partial f_2 \left( \mathbf{q} \right)}{\partial q_u}$ does not exist. Let $q_u = \epsilon$, where $\epsilon$ is a positive infinitesimal, we find the second term of~\eqref{qu_gradient} goes to positive infinity, so~\eqref{qu_gradient} is negative, then we know the first-order derivative is always negative, which means that the optimal solution of $q_u$ is the infinity.

\bibliographystyle{IEEEtran}
\bibliography{ref}
\vfill

\end{document}